\documentstyle[12pt]{article}
\begin{document}
\font\rm=cmr12
\font\tenrm=cmr10
\font\tensl=cmsl10
\font\tenbf=cmbx10
\begin{center}
{}~~~~~
\tenrm
{\tenbf  THEORY OF INCLUSIVE
 SCATTERING OF POLARIZED ELECTRONS BY POLARIZED $^{3}$He
AND THE NEUTRON FORM FACTORS\footnote[1]{Invited talk at VI Workshop on "
Perspectives in Nuclear Physics at Intermediate Energies", ICTP, Trieste,
May 3-7, 1993 (World Scientific, Singapore)}}\\
[0.8cm]
C. CIOFI degli ATTI\\
{\tensl Department of Physics, University of
Perugia, and INFN, Sezione di Perugia,\\
Via A. Pascoli, I-06100 Perugia, Italy}\\[0.5cm]
E. PACE\\
{\tensl Dipartimento di Fisica, Universit\`a di Roma "Tor Vergata",
and INFN, \\
Sezione Tor Vergata, Via E. Carnevale, I-00173 Roma, Italy}
\\[0.3cm]
and
\\[0.3cm]
 G. SALM\`E
\\
{\tensl INFN, Sezione Sanit\`a, Viale Regina Elena 299, I-00161
Roma, Italy}
\\[.8cm]
ABSTRACT\\[0.5cm]
\end{center}
\tenrm
\baselineskip 12pt
\begin{quote}
\rightskip .3cm
\leftskip .3cm

 The theory of inclusive lepton scattering of polarized leptons by polarized
J = 1/2 hadrons is presented and the origin of different expressions for the
polarized nuclear response function appearing in the literature is explained.
The sensitivity of the longitudinal asymmetry upon  the neutron form
factors is investigated.
\end{quote} \vspace{.3cm} \baselineskip 14pt \rm

\noindent{\bf 1. Introduction}
\bigskip

\indent  The scattering of polarized electrons by
polarized targets represents a valuable tool for investigating
nucleon and nuclear properties in great detail$^{1}$. In particular
quasi-elastic (qe) inclusive experiments involving polarized $^3$He $^{2,3}$
are
viewed as possible source of information on the neutron form factors; as a
matter of fact, if only the main component (S-wave)  of the three-body wave
function
is considered, the proton contribution to the asymmetry largely cancels out.
The
proton contribution, arising from
S'  and D-waves, has been studied in Ref.[4] within the closure approximation,
i.e., by describing nuclear effects through spin-dependent momentum
distributions. Adopting the general formalism of Ref.[4], the effects
of nucleon binding has been analysed in Ref.[5], where the concept of the
spin dependent spectral function has been introduced and applied to the
calculation of the $^3$He asymmetry. The effects of  binding has also been
recently considered in Ref.[6], where a new expression for the polarized
nuclear structure functions has been obtained. In that paper, moreover, doubts
have been raised as to the possibility of obtaining reliable information on the
neutron form factors by the measurement of the inclusive asymmetry.

\indent Since  a clear explanation about the origin of the differences
between the expression of the polarized structure functions  used in
Refs.[4,5] and the one obtained in Ref.[6] is lacking in the literature, the
aim
of this paper is first
 to present a comprehensive derivation of the inclusive cross section, in order
to clarify the origin of the above mentioned differences, and,  second,
to show that, provided a proper kinematics is chosen, the qe asymmetry can be
made very sensitive to  neutron properties, and in particular to the neutron
electric form factors.
  \bigskip

\pagestyle{plain}
\noindent{\bf 2. The hadronic tensor and the polarized structure
functions} \bigskip

\indent The inclusive cross section describing the scattering of a
longitudinally
polarized lepton of helicity $h~=~\pm1$ by a polarized hadron of spin J = 1/2,
is given in one photon exchange approximation by $^1$

\begin{eqnarray}
\frac{d^2\sigma(h)}{d\Omega_2 d\nu}~\equiv~
\sigma_2\left(\nu,Q^2,\vec{S}_A,h\right)& =& \frac{4 \alpha^2}{Q^4}\;
\frac{\epsilon_2}{\epsilon_1}\;m^2\;
L^{\mu\nu}W_{\mu\nu}\;=\nonumber\\
&=&\; \frac{4 \alpha^2}{Q^4}\;
\frac{\epsilon_2}{\epsilon_1}\;m^2\;
\left[L^{\mu\nu}_{s}W_{\mu\nu}^{s}+L^{\mu\nu}_{a}
W_{\mu\nu}^{a}\right]
\label{eq1}
\medskip
\end{eqnarray}
where the  symmetric ($s$) and antisymmetric ($a$) leptonic tensors
are

\begin{eqnarray}
 L^{\mu\nu}_{s} &=&\; -(g^{\mu\nu}
+\frac{q^\mu\;q^\nu}{Q^2})\;\frac{Q^2}{4m^2} +~{1 \over m^2}
(k^\mu_1-\frac{q^\mu}{2})\;(k^\nu_1-\frac{q^\nu}{2})
\label{2.1}\\
 L^{\mu\nu}_{a}
&=&\;i~h~\epsilon^{\mu\nu\rho\sigma}\frac{q_\rho\;k_{1\sigma}}{2m^2}
\label{eq2.2} \medskip \end{eqnarray}
and the corresponding  hadronic  tensors are

\begin{eqnarray}
 W_{\mu\nu}^{s} &=&\; -(g_{\mu\nu} +\frac{q_\mu\;q_\nu}{Q^2})\;W^A_1 +
(P_{A\mu}+\frac{P_A\cdot q}{Q^2}q_\mu)\;(P_{A\nu}+\frac{P_A\cdot
q}{Q^2}q_\nu)~\frac{W^A_2}{M^2_A}
\label{3.1} \\
 W_{\mu\nu}^{a}
&=&\;i\epsilon_{\mu\nu\rho\sigma}q^\rho\;V^\sigma
 \label{eq3.2}
\medskip \end{eqnarray}
The pseudovector  $V^\sigma$ appearing in Eq.(\ref{eq3.2}) can be expressed as
follows \begin{equation}
V^\sigma\equiv~S^\sigma_A~\frac{G^A_1}{M_A} +
(P_A\cdot q\;S^\sigma_A - S_A\cdot q\;P^\sigma_A)
{}~\frac{G^A_2}{M^3_A}\label{eq3.3}
\medskip
\end{equation}
In the above equations, the index A
 denotes the number of nucleons  composing
the target, $k^\mu_{1(2)} \equiv (\epsilon_{1(2)},\vec{k}_{1(2)})$
and $P_A^\mu  \equiv (M_A,0)$ are electron and target four-momenta,
$~q^\mu \equiv (\nu,\vec{q})$ is the four-momentum transfer, $Q^2 = -q^2$,
$~g_{\mu\nu}$  is the symmetric metric tensor, $\epsilon_{\mu\nu\rho\sigma}$
the
fully antisymmetric tensor, and $~S_A^\mu$ the polarization four-vector (in the
rest frame $S_A^\mu \equiv (0,\vec{S}_A)$).

\indent Although  both the symmetric,
$W^{s}_{\mu\nu}$, and the antisymmetric, $W^{a}_{\mu\nu}$, parts of the
hadronic tensor are involved in the polarized scattering, in what follows we
will focus on the antisymmetric one, since it contains the relevant
physical quantities we want to investigate. To this end the following remarks
are in order: i) as usual, the general form  of $W^{a}
_{\mu\nu}$, (Eq.(\ref{eq3.2})), can be obtained only by   using invariance
principles (Lorentz, gauge,	parity and time reversal invariance); ii) since the
antisymmetric tensor $\epsilon_{\mu\nu\alpha\beta}$ in Eq.(\ref{eq3.2}) cancels
out  the contribution to $W^{a}_{\mu\nu}$ arising from any term  proportional
to
$q^\mu$ ($\epsilon_{\mu\nu\alpha\beta}~ q^\alpha~q^\beta~=~0$),
terms of this kind, which in principle could appear in the definition of
$V^\sigma$, were not included in Eq.(\ref{eq3.3}). The relevance of this last
comment will be clear later on, when the  method for obtaining  the polarized
structure functions in the framework of PWIA will be discussed.

\indent In total analogy with the case of  the unpolarized structure functions
which are determined by the symmetric part of the hadronic tensor, the
polarized
structure functions  $G_1^A$ and   $G_2^A$ have to be obtained by expressing
them  in terms of the components of  $W^{a}_{\mu\nu}$.
This can be accomplished by  "inverting" Eq.(\ref{eq3.2}). In fact one has

\begin{equation}
\tilde{V}^\sigma~\equiv~V^\sigma~+ {q^\sigma\over Q^2}~(V\cdot q)~=~i {1\over
{2~Q^2}}~\epsilon^{\sigma\alpha\mu\nu}q_\alpha~W^{a}_{\mu\nu}
\label{4}\medskip
\end{equation}
where the four-vector $\tilde{V}^\sigma$ is orthogonal to
$q_\sigma$, i.e. $\tilde{V}\cdot q~=~0$. Thus, working in the rest frame of the
target,  assuming the z-axis
along the momentum transfer  ($\hat{q}~\equiv~\hat{u}_z$), and using
Eq.(\ref{eq3.3}),  the
following expressions for the polarized structure functions $G_1^A$ and
$G_2^A$ are obtained (cf. Ref.[6])

\begin{eqnarray}
{G_1^A\over M_A}&=&-i \left({Q^2\over |\vec{q}|^3}~{W^{a}_{02}\over
S_{Ax}}~+~{\nu\over |\vec{q}|^2} {W^{a}_{12}\over S_{Az}}\right)
\label{5.1}\\
{G_2^A\over M_A^2}&=&-i{1\over |\vec{q}|^2}~\left({\nu\over
|\vec{q}|}{W^{a}_{02}\over S_{Ax}}~-~ {W^{a}_{12}\over S_{Az}}\right)
\label{5.2}\medskip
\end{eqnarray}

\noindent  It should be stressed that, in line with the above  remark ii),
Eqs.(\ref{5.1}) and (\ref{5.2}) are not affected, because of Eq.(\ref{eq3.2}),
by any arbitrary term proportional to $q^\sigma$ which  could be added to
$V^\sigma$ given by Eq.(\ref{eq3.3}). At this point, it should be pointed out
that in Ref.[4] $G_{1(2)}^A$ have been obtained by another procedure, namely by
expressing them in terms of the components of the pseudovector ${V}^\sigma$,
given by Eq.(\ref{eq3.3}). One obtains in this case

\begin{eqnarray}
{G_1^A\over M_A}&=&-{(V\cdot q)\over{|\vec{q}|S_{Az}}}
\label{5.3} \\
{G_2^A\over M_A^2}&=&{V_0\over{|\vec{q}|S_{Az}}}
\label{5.4}
\medskip
\end{eqnarray}
Given the form (\ref{eq3.3}) for $V^\sigma$, Eqs.(\ref{5.1}) and (\ref{5.2})
are totally equivalent to Eqs.(\ref{5.3}) and (\ref{5.4}). However, such an
equivalence will not hold if a term proportional to $q^\mu$ is explicitely
added
to the r.h.s. of Eq.(\ref{eq3.3}), since Eqs.(\ref{5.1}) and (\ref{5.2})
will be not affected by the added term, whereas Eqs.(\ref{5.3}) and
(\ref{5.4}) will be; therefore  $G_1^A$ and  $G_2^A$ obtained from
Eqs.(\ref{5.3}) and
(\ref{5.4}) will be not correct. This remark will be very relevant in the
discussion of the evaluation of $G_1^A$ and  $G_2^A$ in Plane Wave Impulse
Approximation (PWIA) that will be presented in the next Section. To sum up,
unlike the unpolarized case, two different procedures have been followed to
obtain the polarized structure functions $G_{1(2)}^A$; they lead to
Eqs.(\ref{5.1}) and (\ref{5.2}) and Eqs.(\ref{5.3}) and (\ref{5.4}),
respectively; however the latters are correct only in so far as $V^\sigma$ is a
linear combination of only $S^\sigma_A$ and  $P^\sigma_A$ and terms
proportional to $q^\sigma$ are absent in the definition of $V^\sigma$. We will
call the correct prescription leading to Eqs.(\ref{5.1}) and  (\ref{5.2})
prescription I (corresponding to the prescription A of Ref.[6]) and that
leading
to Eqs.(\ref{5.3}) and (\ref{5.4}) prescription II (corresponding  to the
prescription C of Ref.[6] and originally proposed in Ref.[4]).

\bigskip

\noindent{\bf 3. The polarized structure functions in PWIA}
 \bigskip

\indent The equations given in Sect.2 are general ones, relying on the one
photon exchange approximation. When comparing with experimental data, one has
to adopt some models for the nuclear structure functions. All papers so far
published$^{4,5,6}$ use the PWIA. Within such an approximation,
one can obtain the following expression for the antisymmetric hadronic tensor

\begin{equation}
{w}^{a}_{\mu\nu}~=~i\epsilon_{\mu\nu\alpha\beta}~q^\alpha~R^\beta
\label{eq6}
\end{equation}
where the four-pseudovector $R^\beta$ is given by

\begin{eqnarray}
R^\beta~=~\sum\limits_{i=p,n}\left[ {\tilde{G}^i_1(Q^2) \over
M}~\langle~S^\beta~\rangle_i + {\tilde{G}^i_2(Q^2) \over
M^3}~q_\alpha\left
(\langle~p^\alpha~S^\beta\rangle{_i}-\langle~p^\beta~S^\alpha\rangle{_i}\right)\right]
\label{7}
\end{eqnarray}
In Eq.(\ref{7}) $p^\alpha~\equiv~(\sqrt{M^2+{|\vec{p}|}^2},\vec p)$ is the
on-shell nucleon momentum, $\tilde{G}^{p(n)}_{1}(Q^2)$ and
$\tilde{G}^{p(n)}_{2}(Q^2)$ are the proton (neutron) spin-dependent form
factors, related to the Sachs form factors by the following
equations$^4$

\begin{eqnarray}
\tilde{G}^{p(n)}_{1}(Q^2)&=&-~{G_M^{p(n)}\over
2}~{(G_E^{p(n)}+\tau~G_M^{p(n)})\over (1+\tau)} \label{7.1} \\
\tilde{G}^{p(n)}_{2}(Q^2)&=&{G_M^{p(n)}\over
4}~{(G_M^{p(n)}-~G_E^{p(n)})\over (1+\tau)}
\label{7.2}
\medskip
\end{eqnarray}
 with $\tau=Q^2/(4M^2)$, and

\begin{eqnarray}
\langle~(p^\alpha)~S^\beta\rangle{_{p(n)}}~&=&~\int~dE^f_{(A-1)}~\int
d\vec{p}~{M^2\over
E_p~E_{p+q}}~(p^\alpha)~\sum\limits_{l=1,3}~f^{p(n)}_{{\cal
M},l}(\vec{p},E)~S^\beta_l \nonumber \\
&~&\delta(\nu +M_A-\sqrt{(M_{A-1}+E^f_{A-1})^2+{|\vec{p}|}^2}-E_{p+q})
\label{eq7.3}
\medskip
\end{eqnarray}
In
Eq.(\ref{eq7.3})
$E_{p+q}=\sqrt{M^2+(\vec {p}+ \vec{q})^2}$, $E=E^f_{A-1}-E_A$ is the nucleon
removal energy, and $S^\beta_l~\equiv~(\hat{u}_l\cdot \vec{p}/
M,\hat{u}_l+\vec{p}~\hat{u}_l\cdot \vec{p}/( M(E_p +M))~)$ is the polarization
of a moving
nucleon, which in its rest frame has the spin along  the direction of
the $l$-axis, whose versor is $\hat{u}_l$ ($l~=~x,y,z$). Eqs.(\ref{eq6}) and
(\ref{eq7.3}) are a generalization of the expressions of Ref.[4] to the case
where both the nucleon momentum and energy distributions are considered.

\indent In
Eq.(\ref{eq7.3}) the  three-dimensional pseudovector $\vec{f}^{p(n)}_{\cal
M}(\vec{p},E)$ describes the nuclear structure and is  defined as follows

\begin{eqnarray}
\vec{f}^{p(n)}_{\cal
{M}}(\vec{p},E)~=~{\hbox{\large{Tr}}}
 \left(~{\bf\hat{P}}^{p(n)}_{\cal{M}}(\vec{p},E)
{}~\vec{\sigma} \right )
\medskip
\label{eq8}
\end{eqnarray}
where the 2x2 matrix ${\bf\hat{P}}^{p(n)}_{\cal{M}}(\vec{p},E)$ is the spin
dependent spectral function of a nucleon inside a nucleus with  polarization
$\vec{S}_A$ oriented, in general, in a direction different from the  z-axis,
and
$\cal M$ {\em is the  component of the total angular momentum along
$\vec{S}_A$}. The elements of the matrix
${\bf\hat{P}}^{p(n)}_{\cal{M}}(\vec{p},E)$
are  given by
 \begin{eqnarray} P_{\sigma,
\sigma',\cal{M}}^{N} ({\vec{p},E})=\sum\nolimits\limits_{{f}_{A-1}}
{}~_{N}\langle{\vec{p},\sigma;\psi
}_{A-1}^{f} |{\psi }_{J\cal{M}}\rangle~ \langle{\psi
}_{J\cal{M}}|{\psi }_{A-1}^{f};\vec{p},\sigma '\rangle _{N}~
\delta (E-{E}_{A-1}^{f}+{E}_{A})
\label{eq9}
\medskip
\end{eqnarray}
where $|{\psi
}_{J\cal{M}}\rangle$ is the ground state of the target nucleus polarized along
$\vec{S}_A$, $|{\psi }_{A-1}^{f}\rangle$ is an eigenstate of the (A-1) nucleon
system, $|\vec{p},\sigma\rangle_N$ is the plane wave for the nucleon $N\equiv
p(n)$ with the spin along the z-axis equal to $\sigma$. It should be pointed
out
that, for a J = 1/2 nucleus, the trace of ${\bf\hat{P}}^{p(n)}_{\cal{M}}
(\vec{p},E)$ yields the usual unpolarized spectral function$^7$. The spin
dependent spectral function of $^3$He $^5$ has been first obtained from  the
overlap integrals, (Eq.(\ref{eq9})), corresponding to a variational wave
function for the Reid soft-core interaction.  The same quantity has
been calculated in Ref.[6], but using a Faddeev wave function and the Paris
potential.

\indent Since   $\vec{f}^{p(n)}_{\cal{M}}(\vec{p},E)$ is a  pseudovector, it is
a linear combination of the pseudovectors at our disposal, viz. $\vec{S}_A$ and
$\hat{p}~(\hat{p} \cdot \vec{S}_A)$, and therefore it can be put in the
following
form, where any angular dependence is explicitely given,

\begin{eqnarray}
\vec{f}^{p(n)}_{\cal{M}}(\vec{p},E)~=~\vec{S}_A~B_1^{p(n)}(p,E)~+~\hat{p}~(\hat{p}
\cdot \vec{S}_A)~ B_2^{p(n)}(p,E)
\label{eq10}
\medskip
\end{eqnarray}
with $p~\equiv~|\vec{p}|$. The relations between $B_1^{p(n)}$, $B_2^{p(n)}$ and
the quantities ${\hbox{\Large\it P}}_{\parallel}^{p(n)}\!\left( {p} ,
E,\alpha\right)$ and ${\hbox{\Large\it P}}_{\perp}^{p(n)}\!\left( {p} ,
E,\alpha\right)$ used in our previous paper$^5$ can be easily found from
Eqs.(\ref{eq8}), (\ref{eq9}) and (\ref{eq10}) by assuming
$\vec{S}_A~\equiv\hat{q}\equiv$(0,0,1), viz.

 \begin{eqnarray}
{\hbox{\Large\it P}}_{\parallel}^{p(n)}\!\left( {p} , E,\alpha\right)
&=&B^{p(n)}_{1} (p,E)~+~B^{p(n)}_{2} (p,E)~cos^2\alpha
\label{10.1} \\
{\hbox{\Large\it P}}_{\perp}^{p(n)}\!\left( {p},
E,\alpha\right)&=&B^{p(n)}_{2}(p,E)~cos\alpha~sin\alpha
\label{10.2}
\medskip
\end{eqnarray}
 with $cos\alpha~=~\hat{p}\cdot\hat{q}$.

\indent The evaluation of $G^1_A$ and $G^2_A$ in the framework of PWIA can be
carried out by substituting in Eqs.(\ref{5.1}) and (\ref{5.2}) the elements of
the PWIA hadronic tensor $w^{a}_{\mu\nu}$ obtained from
Eqs.(\ref{eq6}), (\ref{7}) and (\ref{eq7.3}), and by using the functions
$B^{p(n)}_{1(2)}$ obtained from Eqs.(\ref{10.1}) and (\ref{10.2}). One gets

\begin{eqnarray}
{}~~~~\frac{G^A_1\left (Q^2,\nu\right )}{M_A}=2\pi
 \sum\nolimits\limits_{i=p,n}
\int \limits_{E_{min}} \limits^{E_{max}(Q^2,\nu)}dE\int
\limits_{p_{min}(Q^2,\nu,E)}\limits^{p_{max}(Q^2,\nu,E)}{p\over{|\vec{q}|E_p}}dp~
\left \{ \tilde{G}_{1}^{(i)}\!\left({Q}^{2}
\right) \left[M~{\hbox{\Large\it
P}}_{\parallel}^{(i)}\!\left( p,E,\alpha\right)~+
\right. \right. \nonumber \\
\left.\left.\;-p~\left({\nu\over|\vec q|}~
-{p~cos\alpha\over{M+E_p}}\right)
{}~\hbox{$\Large\cal P$}^{(i)}\!\left({p}
, E,\alpha\right)\right]~-~{Q^2\over|\vec q|^2}~{\hbox{$\Large\cal
L$}} \right \}~~~~~~~
 \label{eq11}
\end{eqnarray}

\begin{eqnarray}
{}~~~~\frac{G^A_2\left (Q^2,\nu\right )}{M_A^2}=2\pi
  \sum\nolimits\limits_{i=p,n}
\int \limits_{E_{min}} \limits^{E_{max}(Q^2,\nu)}dE\int
\limits_{p_{min}(Q^2,\nu,E)}\limits^{p_{max}(Q^2,\nu,E)}{p\over{|\vec{q}|E_p}}dp~
\left\{ \left[\tilde{G}_{1}^{(i)}\!\left({Q}^{2} \right)~{p\over
|\vec{q}|}{\hbox{$\Large\cal
P$}^{(i)}\!\left({p},E,\alpha\right)}\;+\right.\right.
\nonumber \\
\left.\left.+~{\tilde{G}_{2}^{(i)}\!\left({Q}^{2}
\right)\over M} \left(E_p~{\hbox{\Large\it
P}}_{\parallel}^{(i)}\!\left({p},E,\alpha\right)~-~{p^2~cos\alpha\over{M+E_p}}
{}~\hbox{$\Large\cal P$}^{(i)}\!\left({p} ,
E,\alpha\right)\right)\right]~-~{\nu\over|\vec
q|^2}~{\hbox{$\Large\cal
L$}} \right \}~~~~~~
\label{eq12}
\end{eqnarray}

\noindent with

\begin{eqnarray}
{\hbox{$\Large\cal
L$}}&=&\left[\tilde{G}_1^{(i)}\!\left ( Q^{2} \right)~{\cal
H}_1^i~+~|\vec q|~{\tilde{G}_{2}^{(i)}\!\left ( Q^{2} \right ) \over M}~{\cal
H}_2^i\right ]
\label{12.0}
\\
{\cal H}_1^i&=&{1\over
2}~{(3~cos^2\alpha~-~1)\over
cos\alpha}~\left[{p^2\over{M+E_p}}{\hbox{$\Large\cal P$}^{(i)}\!\left(
{p},E,\alpha\right)}~+~M~{{\hbox{\Large\it
P}}_{\perp}^{(i)}\!\left({p},E,\alpha\right)\over sin\alpha}\right]~~~~~~
\label{12.1}
\\
{\cal H}_2^i&=&p~\left[{\hbox{$\Large\cal
P$}^{(i)}\!\left( {p},E,\alpha\right)}~-~{{\hbox{\Large\it
P}}_{\perp}^{(i)}\!\left({p},E,\alpha\right)\over
sin\alpha}\right]+ \nonumber \\
&~&-~{\nu\over
2|\vec{q}|}~{(3~cos^2\alpha~-~1)\over
cos\alpha}~\left[{p^2\over{M+E_p}}{\hbox{$\Large\cal P$}^{(i)}\!\left(
{p},E,\alpha\right) }~-~E_p~{{\hbox{\Large\it
P}}_{\perp}^{(i)}\!\left({p},E,\alpha\right)\over sin\alpha}\right]
\label{12.2}
\medskip
 \end{eqnarray}
and ${\hbox{$\Large\cal P$}}^{(i)}\!\left( {p},E,\alpha\right) =
cos\alpha~{\hbox{\Large\it
P}}_{\parallel}^{(i)}\!\left({p},E,\alpha\right)~+~sin\alpha~{\hbox{\Large\it
P}}_{\perp}^{(i)}\!\left({p},E,\alpha\right)$. In Eqs.(\ref{eq11}) and
(\ref{eq12})
the integration limits and $cos \alpha$ are determined, as usual, through the
energy conservation$^8$. The polarized structure functions $G^A_{1(2)}$, given
by
Eqs.(\ref{eq11}) and (\ref{eq12}), coincide with the ones corresponding to the
extraction scheme (A) of Ref. [6], once the off-shell effects are neglected,
and
 ${\hbox{\Large\it
P}}_{\parallel(\perp)}^{(i)}\!\left({p},E,\alpha\right)$ are expressed in terms
of the scalar functions $f_1$ and $f_2$ introduced in Ref.[6].

\indent It should be pointed out that $G^A_{1(2)}$, obtained in Ref.[5],
 differ from Eqs.(\ref{eq11}) and (\ref{eq12})
in that they do not contain the term ${\hbox{$\Large\cal
L$}}$. The origin of such a difference is that in Ref.[5] we followed the
procedure of Ref.[4], according to which  the polarized structure
functions are obtained from   Eqs.(\ref{5.3}) and (\ref{5.4}) by directly
substituting the four-vector $V^\sigma$  with the four-vector $R^\sigma$. As
already explained in Sect.2, such a procedure will lead to Eqs.(\ref{eq11}) and
(\ref{eq12}) only if the functional dependence of the two four-vectors is the
same, i.e. if $R^\sigma$ is a linear combination of only $S^\sigma_A$ and
$P^\sigma_A$; it turns out by an explicit evaluation of Eq.(\ref{7})$^9$ that
this is not the case, for $R^\sigma$ contains a term proportional to
$q^\sigma$, which is washed out by  $\epsilon_{\mu\nu\alpha\sigma}$ when the
 hadronic tensor is evaluated.

\bigskip

\noindent{\bf 4. The asymmetry in the quasi-elastic region}
 \bigskip

The contraction of
the two tensors in Eq.(\ref{eq1}) yields

\begin{eqnarray}
\frac{d^2\hbox{$\large\sigma$}(h)}{d\Omega_2 d\nu}~=~
\hbox{$\large\Sigma$}\;+\;h\;\hbox{$\large\Delta$}
\label{eq14}
\end{eqnarray}

where

\begin{eqnarray}
\hbox{$\large\Sigma$}~=~
\hbox{$\large\sigma$}_{Mott} \left
[W^A_2(Q^2,\nu)\;+\;2~tan^2\frac{\theta_{e}}{2}~W^A_1(Q^2,\nu)\right]
\label{eq14.1}
\end{eqnarray}
\begin{eqnarray}
\hbox{$\large\Delta$}~=~
\hbox{$\large\sigma$}_{Mott}~2~tan^2\frac{\theta_{e}}{2}
\left[\frac{G^A_1(Q^2,\nu)}{M_A}~(\vec{k_1}+\vec{k_2})~+~2~
\frac{G^A_2(Q^2,\nu)}{M_A^2}~(\epsilon_1
 \vec{k_2}-\epsilon_2 \vec{k_1})\right]\cdot\vec{S_A}
\label{eq14.2}
\medskip
\end{eqnarray}
In
what follows, the target polarization vector $\vec{S_A}$ is supposed to lie
within the scattering plane formed by $\vec k_1$ and $\vec k_2$.

\indent Two possible kinematical conditions can be considered:

\indent{\em{$\beta$ - kinematics}}. The target polarization angle is measured
with respect to the direction of the incident electron, i.e. $cos\beta =
\vec{S_A} \cdot \vec{k_1}/|\vec{k_1}|$, (this is the more suitable choice from
the experimental point of view). In this case, one gets

\begin{eqnarray}
\hbox{$\large\Delta$}~\equiv~\hbox{$\large\Delta$}_\beta&=&
\hbox{$\large\sigma$}_{Mott}~2~tan^2\frac{\theta_{e}}{2}
\left\{\frac{G^A_1(Q^2,\nu)}{M_A}~\left[\epsilon_1 cos\beta\;+\;\epsilon_2
cos(\theta_{e}-\beta)\right]~+\right. \nonumber \\
&~&\left. -~2~
\frac{G^A_2(Q^2,\nu)}{M_A^2}~\epsilon_1 ~\epsilon_2 \left[cos\beta -
cos(\theta_{e}-\beta)\right]
\right\}
\label{eq16}
\medskip
\end{eqnarray}

\indent{\em{$\theta^*$ - kinematics}}. The target polarization angle is
measured with respect to the direction of the momentum transfer, i.e.
$cos\theta^* = \vec{S_A} \cdot \vec{q}/|\vec{q}|$. Then one can write

\begin{eqnarray}
\hbox{$\large\Delta$}~\equiv~\hbox{$\large\Delta$}_{\theta^*}&=&
-\hbox{$\large\sigma$}_{Mott}~tan\frac{\theta_{e}}{2}
\left\{cos\theta^*~R^A_{T'}(Q^2,\nu)~\left[{Q^2 \over |\vec{q}|^2} +
tan^2\frac{\theta_{e}}{2}\right]^{1/2}~+ \right.
\nonumber \\
&~&\left. -~\frac{Q^2}
{|\vec{q}|^2~\sqrt{2}}~sin\theta^* ~R^A_{TL'}(Q^2,\nu)\right\}
\label{eq17}
\medskip
\end{eqnarray}
where
\begin{eqnarray}
R^A_{T'}(Q^2,\nu)&=& -2~ \left(\frac{G^A_1(Q^2,\nu)}{M_A}~ \nu~ - ~Q^2
\frac{G^A_2(Q^2,\nu)}{M_A^2}\right)
{}~=~i~2~{W^{a}_{12}\over S_{Az}}
 \label{eq17a}
\end{eqnarray}
\begin{eqnarray}
R^A_{TL'}(Q^2,\nu)&=&2~\sqrt{2}~|\vec{q}|~\left(\frac{G^A_1(Q^2,\nu)}{M_A}~
+~\nu
\frac{G^A_2(Q^2,\nu)}{M_A^2}\right)
{}~=~-i~2\sqrt{2}~{W^{a}_{02}\over S_{Ax}}
\label{eq17b}
\end{eqnarray}
In principle the  $\theta^*$ - kinematics is very appealing, since  by
performing experiments at $\theta^* = 0$ and $90^o$ one can disentangle
$R^A_{T'}$ and $R^A_{TL'}$, which, at the top of the qe peak, are proportional
to  $(G^n_M)^2$ and
$G^n_E~G^n_M$, respectively, provided  the proton contribution can be
disregarded$^{2,3}$.

\indent
Experimentally one measures the asymmetry

\begin{eqnarray}
\rm A= \frac {{\sigma }_{2}\left({\nu ,{Q}^{2},{\vec{S}}_{A},+1}\right) -
{\sigma
}_{2}\left({\nu ,{Q}^{2},{\vec{S}}_{A},-1}\right)} {{\sigma }_{2}\left({\nu
,{Q}^{2},{\vec{S}}_{A},+1}\right) + {\sigma }_{2}\left({\nu
,{Q}^{2},{\vec{S}}_{A},-1}\right)} =
\frac{\hbox{$\large\Delta$}}{\hbox{$\large\Sigma$}}
 \label{eq18}
\end{eqnarray}
If the naive model of $^3$He holds, this quantity is in principle very
sensitive
to the neutron properties, since the numerator should be essentially given by
the
neutron with its spin aligned along $\vec S_A$. With this simple picture
in mind, let us consider the comparison between our  results based on
Eqs(\ref{eq11}) and (\ref{eq12}), with the experimental data obtained at
MIT-Bates$^{2,3}$.

\indent In Fig. 1 the asymmetry corresponding to $\epsilon_1 = 574~MeV$ and
$\theta_e=44^o$,   measured  by the
MIT-Caltech collaboration$^2$ is shown. The experimental data were obtained
in a large  interval of the energy transfer after averaging  over three
different values of the $\beta$ angle ( $\beta=44.5^o, 51.5^o,135.5^o$ with the
corresponding azimuthal angles being: $\phi=180^o,180^o,0^o$). It is worth
noting
that in these kinematical conditions one has $\theta^*~\approx~90^o$
only at the top of the qe peak, and  therefore only there the measured
asymmetry reduces to $R_{TL'}$. In the figure  the neutron (dotted line) and
proton (dashed line) contributions are separately  shown, and  the
relevance of the proton  contribution can be seen  particularly at the  top of
the qe peak ($A^{exp}_{qe}\propto~R_{TL'}^{exp}$), where a comparison with the
experimental values (obtained from a further averaging over an  interval of
the energy transfer of about $100~MeV$) yields

\begin{eqnarray}
A^{exp}_{qe}& = & 2.41~\mp1.29~\mp0.51~\%  ~~ MIT-Caltech^2
\nonumber \\
A^{exp}_{qe} & = & 1.75~\mp1.20~\mp0.31~\%  ~~ MIT-Harward^3
\nonumber\\
A^{th}& = & 3.74~\% \nonumber\\
A^{th}_p & = & 2.20~\%
\nonumber
\medskip\end{eqnarray}

\indent In Fig.2 the theoretical asymmetry, averaged over the same values of
the
polarization angle of the previous case, is shown in correspondence with
$\epsilon_1~=574~MeV$ and   $\theta_e~=~51.1^o$ . Such a kinematics was
chosen$^{2,3}$ with the aim of extracting $R_{T'}$ at the qe peak. Only one
experimental point has been obtained  for the averaged asymmetry around the
top of the qe peak, where $\theta^*~\approx~0^o$
($A^{exp}_{qe}\propto~R_{T'}^{exp}$). The comparison between the experimental
results and our calculation is as follows

\begin{eqnarray}
A^{exp}_{qe}&=&-3.79~\mp1.37~\mp0.67~\%~~MIT-Caltech^2
\nonumber\\
A^{exp}_{qe}&=&-2.60~\mp0.90~\mp0.46~\%~~MIT-Harward^3
\nonumber\\
A^{th}&=&-3.43~\% \nonumber\\
A^{th}_p&=&-1.30~\%
\nonumber
\medskip\end{eqnarray}
It should be pointed out that our numerical results, as shown in Fig. 3,  are
only slightly different from the ones obtained in Ref.[6], where a
spin-dependent Faddeev spectral function has
been used.

\indent In Fig. 4a and 4b  the results based upon prescription
I are compared with the our  previous one$^5$, based upon prescription
II (the corresponding explicit expressions for $G^A_{1(2)}$ are given in
Ref.[5]
and coincide, as already mentioned, with Eqs.(\ref{eq11}) and (\ref{eq12}) with
the term ${\hbox{$\Large\cal L$}}$  dropped out). The results of the comparison
(cf. Fig. 4a and 4b) show that at $\theta^*~\approx~90^o$ the approximate
method yields results very different from the correct one, whereas at
$\theta^*~\approx~0^o$
 such a difference is not present.   This is due to the fact that  in the
procedure II $R_{TL'}$ results to be proportional to the component of $\vec{R}$
along $\hat{q}$, instead of being proportional to its transverse part, as it
should be$^9$, therefore it is affected by the difference between $V^\sigma$
and
$R^\sigma$, arising from the term
 proportional to $q^\sigma$ which is present in the latter.  For
$\theta^*~\approx~0^o$ the differences, as shown in Fig.4b, are very small over
the whole range of the energy transfer considered, since $R_{T'}$ is unaffected
by the extra term in$~R^\sigma$ ( we recall that $A~\propto~R_{T'}$ only at the
top of the qe peak, and the mixing with $R_{TL'}$ explains the small
differences
on the wings of the asymmetry).

\noindent From the above comparisons it turns out  that the difference between
the two procedures is almost entirely due to the proton contribution; for such
a
reason the correctness of our conclusion, reached in [5] using prescription II,
about the possibility of obtaining information on the neutron form factors by
properly minimizing the proton contribution are not affected by the use of
prescription I. This will be illustrated in the next Section.

\begin{figure}

\parbox{5.5cm}{{\tenrm	Fig. 1.  The asymmetry corresponding to  $\epsilon_1$ =
574~$MeV$ and $\theta_e~ = ~44^o$,  vs. the
energy transfer $\nu$ calculated by  Eqs. (22) and (23) (solid line) and using
the spin-dependent spectral function of Ref.[5]; the dotted (dashed) line
represents the neutron (proton) contribution. The nucleon
form factors of Ref. [10] have been used and the experimental data are from
Ref.[2]. The arrow indicates the position of the qe peak.}

\vspace{3cm}
{\tenrm Fig. 2 The same as in Fig. 2, but for $\theta_e~ =~ 51.1^o$. The
experimental point, has been obtained , Ref.[2], after averaging over a 103 MeV
interval around the qe peak, as explained in the text.}

\vspace{3cm}
{\tenrm	Fig. 3. Comparison of the asymmetry ( $\epsilon_1$ =
574~$MeV$ and $\theta_e~ = ~44^o$) shown in Fig. 1 ( calculated by  Eqs. (22)
and
(23)) (solid line) with the
one of Ref.[6] (dashed line), based on a Faddeev spin-dependent spectral
function. The  nucleon form factors of Ref. [11] have been used and the
experimental data are from
Ref.[2]. The arrow indicates the position of the qe peak. }}
 \ $~~$ \
\parbox{9cm}{\raisebox{-22cm}{\special{picture trst123b}}}
\end{figure}
\begin{figure}

\parbox{5.5cm}{
{\tenrm	Fig. 4a. The asymmetry and the neutron contribution  for $\epsilon_1$ =
574 $MeV$ and $\theta_e~ =~ 44^o$   vs. the
energy transfer $\nu$ calculated using prescriptions I and II. Solid (dotted)
line: the
asymmetry (neutron contribution) corresponding to prescription I ( Eqs. (22)
and
(23));  dashed (dot-dashed)
line: the asymmetry (neutron contribution) corresponding to
 prescription II ( Eqs. (22) and
(23) without the term $\cal L$). The form factors of Ref.
[10] have been used and the experimental data are from
Ref.[2]. The  arrow indicates the position of qe peak.}
\vspace{2cm}

{\tenrm Fig. 4b. The same as in Fig. 4a, but for  $\theta_e~ =~
51.1^o$.}
\vspace{5cm}

{\tenrm	Fig. 5. The proton  contribution
to the asymmetry , at the top of the qe peak , vs. $\beta$, for
$\theta_e~=~75^o$. Solid line: $\epsilon_1~=~500~MeV$, long-dashed line:
$\epsilon_1~=~1000~MeV$, short-dashed line: $\epsilon_1~=~1500~MeV$, dotted
line:
$\epsilon_1~=~2000~MeV$. The nucleon form factors of Ref. [10] have been used.}
\vspace{1cm}
} \ $~~$ \
\parbox{9cm}{\raisebox{-22cm}{\special{picture trst445}}} \end{figure}

\bigskip
\noindent{\bf 5. Minimizing the proton contribution}
\bigskip

\indent As shown in Figs. 1 and 2, the proton contribution to the asymmetry,
corresponding to the polarization angle of the actual experiments is sizeable.
Following our previous paper$^5$, in this Section  the possibility to minimize
or
even to make vanishing the proton contribution will be investigated. To this
end
we have analyzed the proton contribution to the asymmetry at the top of the qe
peak,  for different values of $\beta$, different values of the energy of the
incident electron and different models for the proton form factors. The results
are presented in Fig. 5. It can be seen that the proton contribution is almost
vanishing around $\beta\equiv\beta_c=95^o$, in a large spectrum of values of
incident electron energy; such a feature, moreover, weakly depends upon the
model for the nucleon form factors.
 In Fig. 6,  the
asymmetry, and the proton contribution, vs of $Q^2$, at fixed values of
$\beta_c~=~95^o$ and $\theta_e~=~75^o$ is presented for three different models
of the nucleon form factors (Refs.[10, 12-13]). It can be seen that the
asymmetry
is very sensitive to the neutron form factors, but from Fig. 6 it is not
possible
to assess whether the differences in the asymmetry are given by the differences
in   $G_E^n$ or in $G^n_M$, since both of them vary within the  models we have
considered. In order to make our analysis a more stringent one, we have
repeated
the calculation by using the Galster model of the nucleon form factors$^{11}$,
since  within such a model  $G_E^n$ can be changed independently of $G_M^n$. In
fact one has
 \begin{figure}
 \parbox{5.5cm}{
{\tenrm	Fig. 6. The total asymmetry  at the top of the qe
peak, vs. Q$^2$, for  $\theta_e~=~75^0$ and $\beta~=~95^o$, using Eqs. (22) and
(23). Solid line : Gari-Kruempelmann form factors$^{10}$; dashed line:
Blatnik-Zovko form factors$^{13}$, dotted line: Hoehler et al form
factors$^{12}$. The curves in the lower part of the figure represent the
corresponding proton contributions.}} \ $~~$ \
\parbox{9cm} {\raisebox{-6cm}{\special{picture trst6}}}
\end{figure}

\begin{figure}
 \parbox{5.5cm}{
{\tenrm	Fig. 7.  The same as in Fig.
6, but for the Galster form factors [11].}} \ $~~$ \
\parbox{9cm} {\raisebox{-6cm}{\special{picture trst70}}}
\end{figure}

\begin{eqnarray}
G_M^n&=&{\mu_n~G_E^p} \nonumber \\
G_E^n&=&{-\tau~\mu_n\over(1+\eta~\tau)}~G_E^p
\label{19}
\medskip
\end{eqnarray}
where  $G_E^p = 1/(1+Q^2/B)^2$, $B~=~0.71 (GeV/c)^2$ and $\eta$ is a parameter.
The resulting asymmetry and proton contribution are shown in Fig. 7 for
different values of  $\eta$.  Fig.7
illustrates  how the total asymmetry can depend upon  $G^n_E$, having
a vanishing proton contribution.

\noindent It should be stressed that the
proposed kinematics, which minimizes the proton contribution, corresponds to
the  qe peak, where the final state interaction is expected to play a minor
role.

\bigskip
\noindent{\bf 6. Summary and conclusion}
\bigskip

\indent The qe spin-dependent structure functions for a nucleus with J =1/2
have
been obtained by a proper procedure, based on the replacement of the exact
hadronic tensor with its PWIA version. Our formal results are in agreement with
the ones of Ref.[6], whereas the numerical calculations only slightly differ,
which demonstrates the equivalence of the spin dependent spectral functions
used
in Ref.[5] and  Ref.[6].

\indent The origin of the differences between the predictions of the correct
procedure and the ones$^{4,5}$ based upon the replacement of the hadronic
pseudovector $V^\sigma$$^4$ with its PWIA version $R^\sigma$,
Eq.(\ref{eq3.3}), have been clarified. In particular it has been shown that
these differences are produced by  the presence of an extra term proportional
to
the momentum transfer $q^\sigma$ in the four-vector $R^\sigma$, Eq.(\ref{7}).
This extra term affects only the response function $R_{TL'}$.

\indent Our analysis of the asymmetry, based on the correct expression
of $G^A_1$ and $G^A_2$ given by Eqs.(\ref{eq11}) and
(\ref{eq12}) respectively, has fully confirmed the main conclusions of our
previous paper$^5$, concerning : i) the relevance of the proton contribution
for
the experimental kinematics  considered till now, and ii) the possibility of
selecting a polarization angle, which leads at qe peak to an almost vanishing
proton contribution for a wide range of the kinematical variable; within such a
kinematical condition, the sensitivity of the asymmetry to the electric
neutron form factor has been thoroughly investigated.

\indent Calculations of the final state effects are in progress.

\bigskip
\noindent{\bf 6. References}
\medskip
\begin{itemize}
\item[ 1.]	T. W. Donnelly, A. S. Raskin, {\em Ann. Phys. (N.Y.)} {\bf 169}
(1986) 247.
\item[2.] a) C. E. Woodward
et al., {\em{Phys. Rev. Lett.}} {\bf{65}} (1990) 698; b) C. E. Jones-Woodward
et
al., {\em{Phys. Rev.}} {\bf{C}} 44 (1991) R571; c) C. E. Jones-Woodward et
al., {\em{Phys. Rev.}} {\bf{C}} 47 (1993) 110.

R. Milner this workshop
\item[3.] A. K. Thompson et
al, {\em{Phys. Rev. Lett.}} {\bf{68}} (1992) 2901; A. M. Bernstein,
{\em{Few-Body Systems Suppl.}} {\bf{6}}  (1992) 485.
\item[4.] B.
Blankleider and R.M. Woloshyn, {\em{Phys. Rev.}} {\bf{C 29}}  (1984) 538.
\item[5.] C. Ciofi degli Atti, E. Pace and G.
Salm\`e,{\em{Phys. Rev.}} {\bf{C 46}}  (1992) R1591.
\item[6.] R.W. Schultze and P.U. Sauer {\em{Phys. Rev.}} {\bf{C 48}} (1993)
xxx.
 \item[7.] C. Ciofi degli Atti, E. Pace and G.
Salm\`e, {\em Phys. Lett.} {\bf B141} (1984) 14.
\item[8.] C. Ciofi degli Atti, E. Pace and G.
Salm\`e, {\em{Phys. Rev.}} {\bf{C 43}}  (1991) 1155.
\item[9.] C. Ciofi degli Atti, E. Pace and G.
Salm\`e, to be published.
\item[10.] M. Gari and W. Krumpelmann, {\em Z. Phys.} {\bf A322} (1985) 689;
{\em Phys. Lett.} {\bf B 173} (1986) 10.
\item[11.] S. Galster et al, {\em Nucl. Phys.} {\bf B32} (1971) 221.
\item[12.]	G. Hoehler et al., {\em Nucl. Phys.} {\bf B 114} (1976) 505.
\item[13.]	S. Blatnik and N. Zovko, {\em Acta Phys. Austriaca} {\bf 39} (1974)
62.
\end{itemize}
\end{document}